Article

# Interchange reconnection as the source of the fast solar wind within coronal holes

S. D. Bale[1,2 ✉], J. F. Drake[3,4], M. D. McManus[1,2], M. I. Desai[5], S. T. Badman[6], D. E. Larson[2], M. Swisdak[4], T. S. Horbury[7], N. E. Raouafi[8], T. Phan[2], M. Velli[9,10], D. J. McComas[11], C. M. S. Cohen[12], D. Mitchell[8], O. Panasenco[13] & J. C. Kasper[14,15]

The fast solar wind that fills the heliosphere originates from deep within regions of open magnetic field on the Sun called 'coronal holes'. The energy source responsible for accelerating the plasma is widely debated; however, there is evidence that it is ultimately magnetic in nature, with candidate mechanisms including wave heating[1,2] and interchange reconnection[3–5]. The coronal magnetic field near the solar surface is structured on scales associated with 'supergranulation' convection cells, whereby descending flows create intense fields. The energy density in these 'network' magnetic field bundles is a candidate energy source for the wind. Here we report measurements of fast solar wind streams from the Parker Solar Probe (PSP) spacecraft[6] that provide strong evidence for the interchange reconnection mechanism. We show that the supergranulation structure at the coronal base remains imprinted in the near-Sun solar wind, resulting in asymmetric patches of magnetic 'switchbacks'[7,8] and bursty wind streams with power-law-like energetic ion spectra to beyond 100 keV. Computer simulations of interchange reconnection support key features of the observations, including the ion spectra. Important characteristics of interchange reconnection in the low corona are inferred from the data, including that the reconnection is collisionless and that the energy release rate is sufficient to power the fast wind. In this scenario, magnetic reconnection is continuous and the wind is driven by both the resulting plasma pressure and the radial Alfvénic flow bursts.

Recent measurements from the NASA Parker Solar Probe (PSP) showed that the solar wind emerging from coronal holes is organized into 'microstreams' with an angular scale (5–10°) in the Carrington longitude[9] similar to the underlying supergranulation cells associated with horizontal flows in the photosphere[10]. However, the footpoints of the previous PSP encounter were at high latitudes on the far side of the Sun, so that the magnetic structure of the cells and their connectivity to the spacecraft could not be determined, preventing a complete analysis of the source of the microstreams.

On solar Encounter 10 (E10), PSP came within 12.3 solar radii ($R_S$) of the photosphere. Figure 1 summarizes the plasma[11], energetic ion[12] and magnetic field measurements[13] made near perihelion. An ion spectrogram in Fig. 1a,b extends from thermal energies to around 85 keV and, like the proton velocity in Fig. 1c, is structured as discrete 'microstreams'[9,14,15] whose duration decreases from around 10 h to around 2 h as the spacecraft approaches perihelion. Data in Fig. 4b (and discussed later) show that the ion energy distributions are power laws at high energy that extend to beyond 100 keV. The characteristic structure of the microstreams is highlighted by red arcs in Fig. 1c, and a blue trace indicates the measured thermal alpha particle abundance $A_{He} = n_\alpha/n_p$ (where $n_\alpha$ and $n_p$ are the alpha particle density and proton number density, respectively), which is similarly modulated. The high first ionization potential of helium requires that the alpha particle abundance is frozen-in at the base of the corona or in the chromosphere[16], so these microstream structures are organized at the source of the wind itself. The radial component of the interplanetary magnetic field in Fig. 1d shows that large-amplitude, Alfvénic field reversals, 'switchbacks', are also associated with the microstreams. A potential field source surface (PFSS) model[17–19] (Methods) is used to infer the footpoints of the magnetic field that connects to the PSP and shows connection to two distinct coronal holes. The time series of the longitude of the footpoint on the solar surface is shown in Fig. 1e and as white diamonds against a 193-Å Solar Dynamics Observatory/Extreme Ultraviolet[20] image in Fig. 2a.

Correspondence of the temporal structure of the switchback and radial velocity bursts with the spatial periodicity of the surface magnetic field documented in Figs. 1 and 2 suggests that magnetic reconnection between open and closed magnetic fields in the low

[1]Physics Department, University of California, Berkeley, CA, USA. [2]Space Sciences Laboratory, University of California, Berkeley, CA, USA. [3]Department of Physics, the Institute for Physical Science and Technology and the Joint Space Institute, University of Maryland, College Park, MD, USA. [4]Institute for Research in Electronics and Applied Physics, University of Maryland, College Park, MD, USA. [5]Southwest Research Institute, San Antonio, TX, USA. [6]Harvard-Smithsonian Center for Astrophysics, Cambridge, MA, USA. [7]The Blackett Laboratory, Imperial College London, London, UK. [8]Johns Hopkins Applied Physics Laboratory, Laurel, MD, USA. [9]Department of Earth, Planetary, and Space Sciences, University of California, Los Angeles, CA, USA. [10]International Space Science Institute, Bern, Switzerland. [11]Department of Astrophysical Sciences, Princeton University, Princeton, NJ, USA. [12]California Institute of Technology, Pasadena, CA, USA. [13]Advanced Heliophysics Inc., Los Angeles, CA, USA. [14]BWX Technologies, Inc., Washington, DC, USA. [15]Climate and Space Sciences and Engineering, University of Michigan, Ann Arbor, MI, USA. ✉e-mail: bale@berkeley.edu

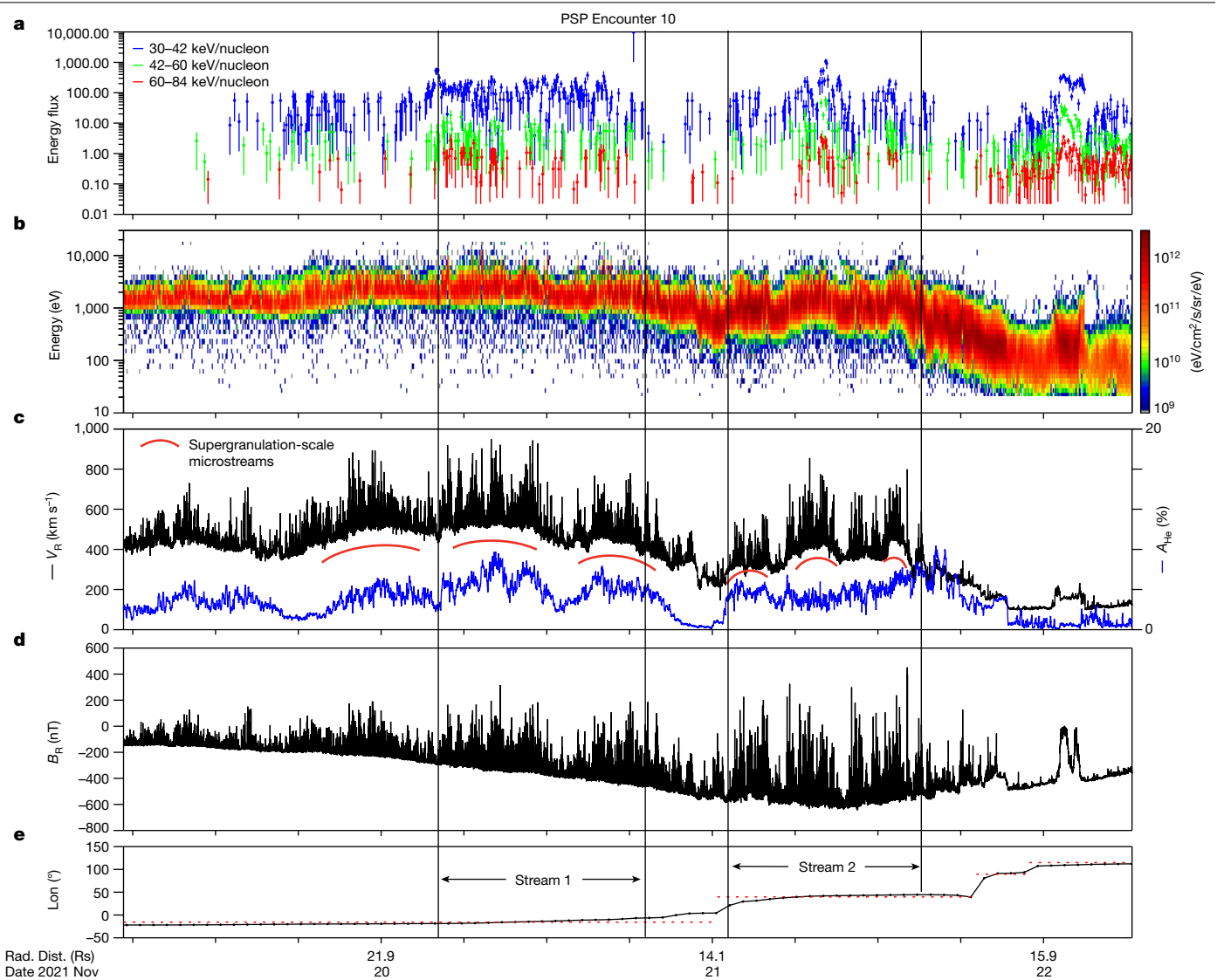

**Fig. 1 | Time series measurements of the solar wind plasma and magnetic field through the November 2021 solar encounter. a,b**, Hot solar wind ions in **a** extend in energy to greater than 85 keV as suprathermal tails on the proton particle distribution in **b**. **c**, Red arcs mark the solar wind radial velocity ($V_R$) microstream structure that is organized in Carrington longitude at angular scales associated with supergranulation convection and the photospheric network magnetic field (Fig. 2). These microstreams become shorter in duration as the spacecraft accelerates through perihelion near the centre of this figure and sweeps more rapidly through Carrington longitude. The thermal alpha particle abundance ($A_{He}$, blue trace in **c**) is similarly modulated by the microstream structure. The alpha particle abundance is frozen-in at the base of the corona. **d**, Reversals of the radial magnetic field ($B_R$), so-called 'switchbacks', are organized by the microstreams and are linked to the radial flow bursts by the Alfvénicity condition. **e**, Photospheric footpoints from a PFSS model instantiation indicate two distinct coronal hole sources well separated in Carrington longitude (Lon), shown in Fig. 2 (and as dotted lines in Fig. 1e).

corona (interchange reconnection) is the driver of these bursts[9,21–23]. Interchange reconnection in the weakly collisional corona is expected to be bursty rather than steady[24–27]. The energetic ions and enhanced pressure in these bursts are also signatures of reconnection[28–30]. The data suggest that it is a continuous process in the source regions of open flux. Figure 3c is a schematic that shows open flux reconnecting with closed flux regions in the low corona. In this figure the open flux migrates to the left, reconnecting with successive regions of closed flux, with the consequence that the bursty outflow from interchange reconnection fills all of the open flux, as seen in the data.

To establish that interchange reconnection is the source of the bursty flows, we use the measurements and established principles of reconnection to deduce the basic characteristics in the low corona. The strength of the reconnecting magnetic field is a key parameter. As the field strength at the base of the corona has substantial variation, we estimate the amplitude of the reconnecting magnetic field by projecting the measured magnetic field at the PSP back to the solar surface. The $R^{-2}$ fall-off of the radial magnetic field with heliospheric distance $R$ is valid in the solar wind, but fails closer to the Sun. Thus, we use a combination of the $R^{-2}$ behaviour at large $R$ with a fall-off derived from a surface-averaged PFSS model below $2.5\,R_s$ (Extended Data Fig. 1). The resulting projection of the 600-nT magnetic field at $13.4\,R_s$ to the low corona is 4.5 G, which is consistent with the PFSS data in Fig. 2. Plasma density at the base of the corona is not measured directly. However, the characteristic amplitude of the bursty flows at PSP are around 300 km s$^{-1}$. Because the flows during bursty reconnection are Alfvénic, we can estimate the density knowing the magnetic field strength. The resulting density is around $10^9$ cm$^{-3}$, a reasonable value for the low corona[31].

# Article

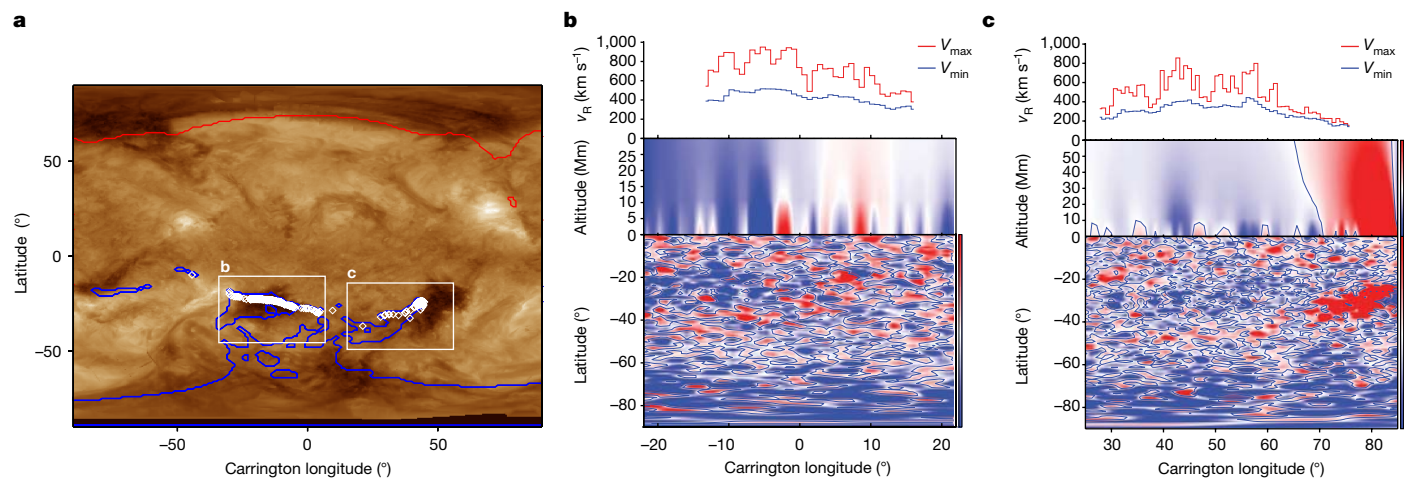

**Fig. 2 | Solar wind during PSP Encounter 10 emerges from two coronal holes. a**, An extreme ultraviolet (193 Å) map of the corona shows cooler regions (darker pixels) associated with open magnetic field within two separate, near-equatorial coronal holes. A PFSS model maps the interplanetary magnetic field from the PSP spacecraft to footpoints (white diamonds) within the coronal holes. **b**, The magnetic field and velocity microstream profile within the first coronal hole: the upper panel shows minimum (blue) and maximum (red) radial speed versus longitude, and the second panel shows the vertical magnetic field along the footpoints extending from the photosphere to 30 Mm from magnetogram measurements and a PFSS model that accounts for the motion of the spacecraft. The bottom panel is a map of the magnetic field polarity just above the photosphere, again from the PFSS model. **c**, The corresponding structure within the second coronal hole. These data indicate that the radial magnetic field is organized into mixed radial polarity intervals on the same scales as the velocity microstreams observed by PSP.

To address whether the rate of energy release is sufficient to drive, the wind we estimate the reconnection inflow rate $V_r$. A lower limit follows from the fact that the flow bursts are nearly continuous. We define the reconnection time as $t_r = L_B/V_r$, the time required for open field lines to traverse the characteristic scale length $L_B$ of the surface magnetic field, which is around 10° or $6 \times 10^4$ km. A second time is the time $t_b \approx R_{PSP}/V_R$ for the reconnection bursts to reach the spacecraft at $R_{PSP}$. In the limit $t_r \gg t_b$, the outflows from the reconnection site would quickly pass by the spacecraft and there would be no high-speed flows until the spacecraft connected to another reconnection site. When $t_r \leq t_b$, the spacecraft would measure bursty flows as the spacecraft crossed the entire supergranulation scale. The observations show the latter because bursty flows are measured during the entire crossing of the supergranulation scale. Observations suggest that $t_r \leq t_b$ or $V_r \approx L_B V_R/R_{PSP}$ is ≈3 km s$^{-1}$ or around 0.01 of the local Alfvén speed, a low value if reconnection is collisionless,[32–34] but comparable to the magnetohydrodynamic (MHD) prediction[35]. With ambient temperatures of around 100 eV, the reconnection electric field is therefore around three orders of magnitude above the Dreicer runaway field. In this regime, collisions are too weak to limit electron acceleration, and collisionless processes dominate. The rate of magnetic energy release from interchange reconnection is given by $V_r B^2/4\pi \approx 5 \times 10^5$ ergs cm$^{-2}$ s$^{-1}$ using $B = 4.5$ G and $V_r = 3$ km s$^{-1}$. This is comparable to that required to drive the high-speed wind, which is around $10^5$–$10^6$ ergs cm$^{-2}$ s$^{-1}$.

Thus through the PSP observations, the Solar Dynamics Observatory/Helioseismic and Magnetic Imager (SDO/HMI) surface magnetic field measurements and well-known characteristics of magnetic reconnection, we have established that interchange reconnection is sufficient to drive both the ambient base solar wind flow through the radial pressure drop and the microstream bursts that lie on top of this flow. Further tests of the reconnection scenario concern the structuring of the flow bursts and the production of energetic protons and alphas. A key observation reported in the E06 data[9] and illustrated in the schematic in Fig. 3c is the temporal asymmetry in the bursts: large-amplitude bursts onset sharply and decrease gradually across the burst period and the time sequence then repeats. Data from a particle-in-cell (PIC) simulation is presented in Fig. 3b (Methods). A cut across the simulation outflow exhaust reveals high-speed bursts on newly reconnected field lines in the exhaust adjacent to the magnetic separatrix, whereas, on field lines in the exhaust interior, the fastest flow bursts have already passed the location of the cut, so the measured flows are weaker (Fig. 3a). Our simulations support the hypothesis that the bursts observed by the PSP correspond to crossings of interchange reconnection exhausts. Dispersion signatures are well documented in the cusp of the Earth's magnetosphere as a result of reconnection at the terrestrial magnetopause[36]. Reconnection between the closed magnetic flux of the Earth and 'open' flux in the solar wind is an analogue of coronal interchange reconnection.

Finally, the spectrum of energetic protons and alphas has been calculated from the interchange reconnection simulations. The simulation includes fully stripped alpha particles that are 5% by number, similar to the solar atmosphere[16]. Energy flux spectra of both species are shown in Fig. 4a. Data is taken from the outflow exhaust and includes only plasma that has undergone acceleration. Protons and alphas exhibit an energetic, non-thermal power-law distribution with spectral indices of around −8 for both species. As shown in Fig. 4b, in the spectrum of the differential energy flux of particles during the time 04:00–19:00 on 20 November 2021 (from Fig. 1), there are also energetic protons and alphas with energies beyond 100 keV. The spectra are again rather soft, having spectral indices of around −9, consistent with the simulation data. The energy in the simulation is normalized to the free parameter $m_i V_A^2$. By equating the energy minimum of the proton power law in the simulation (approximately $5\, m_i V_A^2$) to that of the PSP measurements (approximately 7 keV), we find that the coronal value of $m_i V_A^2$ is around 1.4 keV, compared with around 0.9 keV from the 300 km s$^{-1}$ estimate for $V_A$ based on the amplitude of the bursty flows measured at 13.4 $R_s$. That the two values of $m_i V_A^2$ are close indicates that the Alfvén speed in the corona where reconnection is taking place is in the range of 300–400 km s$^{-1}$.

The picture that emerges is that reconnection directly heats the ambient coronal plasma sufficiently to drive the bulk outflow[37–39] and at the same time produces the turbulent velocity bursts that ride this outflow[24–27]. Extended Data Fig. 2 shows the strong heating of protons from the simulation in Fig. 3a,b. Of course, a fraction of the magnetic energy released during reconnection can take the form of Alfvén waves[1,2] or other magnetic structures[5] that can be dissipated

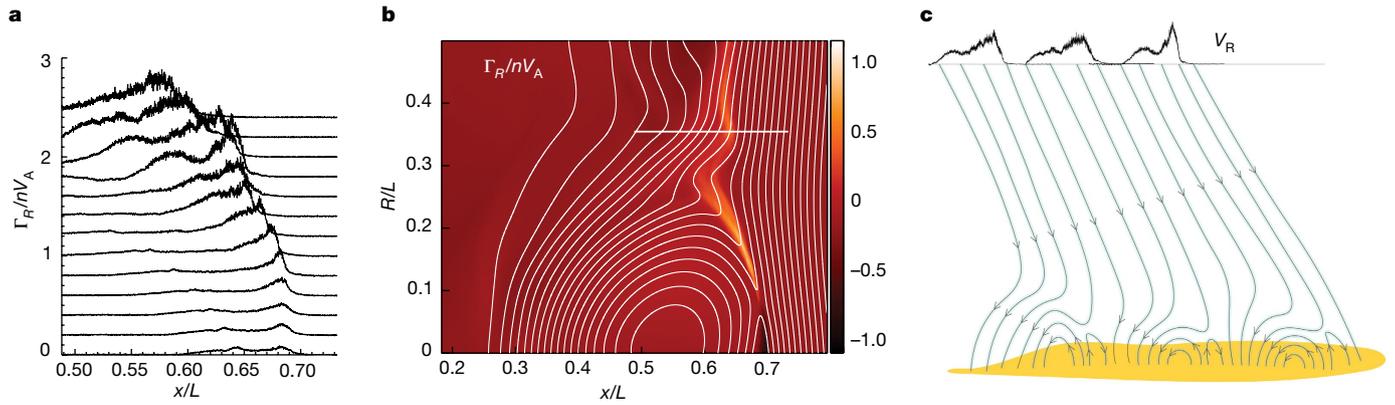

**Fig. 3 | Schematic of interchange reconnection and the structure of the reconnection exhaust from a simulation. a**, A time sequence of the dimensionless radial flux along the cut shown by the horizontal white line in **b**. Each successive cut is separated by a time $0.037\, L/V_A$, where $L/V_A$ is the Alfvén transit time across the simulation domain, and is shifted upward to avoid overlapping the data. The cuts reveal the bursty nature of the outflow resulting from the generation of flux ropes within the elongated current layer[24–27]. As shown in the schematic, newly reconnected field lines have higher outflow fluxes than field lines reconnected earlier in time. **b**, The dimensionless radial flux with overlying magnetic fields in white from a PIC simulation of interchange reconnection showing Alfvénic upward and downward flows from the reconnection site above the coronal surface. Details about the simulation set-up are found in the Supplementary material. The reconnected magnetic field migrates to the left as it straightens and drives the outflow exhaust. **c**, A schematic of reconnection between open and closed magnetic flux (interchange reconnection) in the low corona based on the PSP data shown in Fig. 1. The data suggest that reconnection between open and closed flux is nearly continuous. In the schematic, the open magnetic flux is moving continuously to the left. An open field line first reconnects with the closed flux above the solar surface, forming upward and downward oriented loops. The open bent field then straightens and drives Alfvénic flow outward. As it moves to the left, the open field line then intersects another closed flux region and the process repeats. Thus, the open flux is completely filled with high-speed outflowing plasma – the exhaust from interchange reconnection. The cuts of the radial velocity measured by an observer crossing the open flux at the top of the schematic indicate that the highest-speed bursty outflows are on newly reconnected magnetic fields whereas, on field lines that reconnected earlier, the highest-speed flows have already passed by the observation location. This time asymmetry was clear in the E06 data[9].

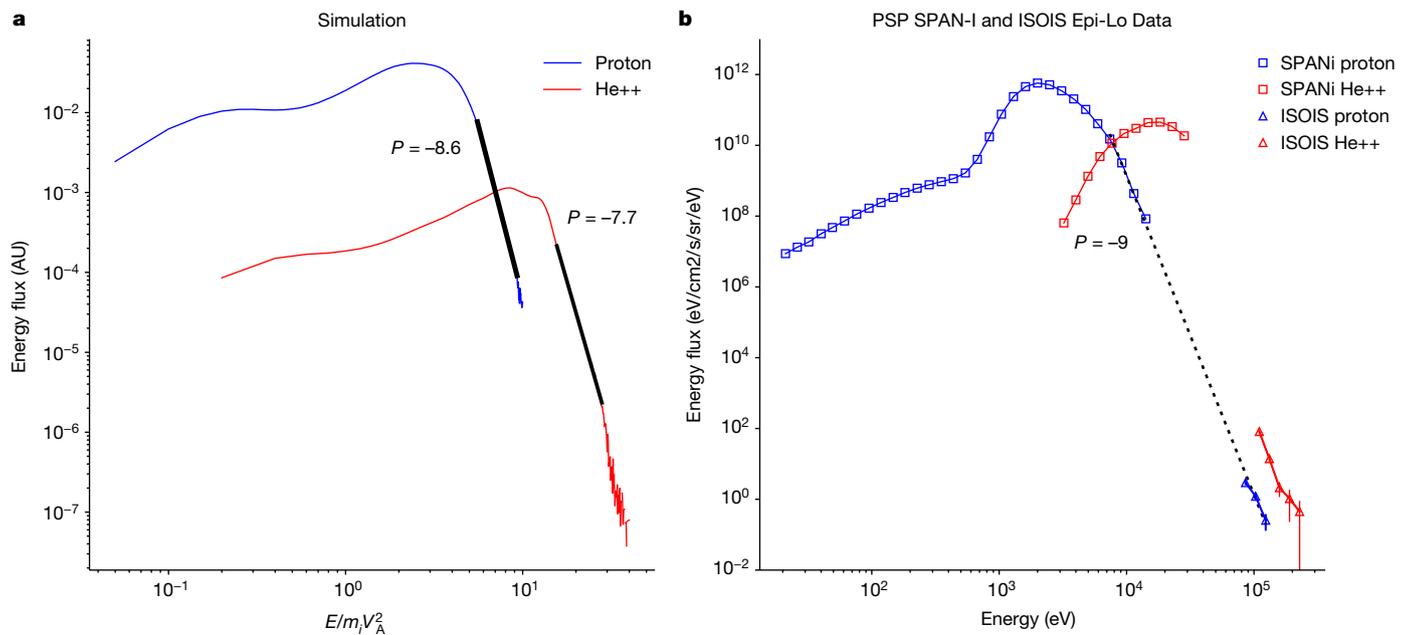

**Fig. 4 | Proton and alpha particle differential energy flux from interchange reconnection simulations and solar wind observations. a**, The proton (blue) and alpha particle (red) energy fluxes taken from the outflow exhaust from an interchange reconnection simulation (see Methods for simulation details). The energy normalization in the simulation is $m_i V_A^2$, which is an arbitrary parameter[35]. The units in the ordinate direction are arbitrary, although the reduced height of the alpha flux reflects the 5% number density of alphas. Both fluxes peak and then roll over into distinct soft power laws with slopes of −8.6 and −7.7 for the protons and alphas, respectively, with the alpha spectrum shifted to higher energy than that of the protons. The spectral indices of the energetic ions depend on the magnitude of the ambient guide (out-of-plane) magnetic field with stronger guide fields producing softer spectra. The data are from a simulation with a guide field of 0.55 of the reconnecting magnetic field. **b**, The proton (blue) and alpha (red) energy fluxes from PSP measurements during the time interval 04:00:00–19:00:00 on 20 November 2021 from Fig. 1. As in the simulations, the spectra peak and roll over into power-law-like suprathermal tails with a similar slope of −9 for the protons. Thus, the power-law slopes from the simulation and the observational data are very close. The alpha measurement does not extend to high enough energy to characterize any power-law behaviour. Finally, we can use the low energy bound of the power-law distribution from the simulation (approximately $5\, m_i V_A^2$) and observations (approximately 7 keV) to establish that the value of $m_i V_A^2$ at the coronal reconnection site is around 1.4 keV. This is comparable with around 0.9 keV from the 300 km s$^{-1}$ estimate for $V_A$ based on the amplitude of the bursty flows measured by the PSP at 13.4 $R_s$.

# Article

higher in the coronal to further drive the bulk outflow[1,2,5,37]. However, the in situ data from reconnection in the Earth's magnetosphere[40] and at the heliospheric current sheet[41] show strong local plasma energization rather than wave generation. The time asymmetry that characterizes the bursty flows[9] and the spectral indices of the power-law distributions of energetic ions are in remarkably good agreement with the interchange reconnection simulation data, in which local plasma energization dominates waves and turbulence. Still, three-dimensional simulations with greater scale separation might show stronger magnetic turbulence. In either scenario, interchange reconnection is the likely energy drive mechanism of the fast solar wind. Recent remote-sensing measurements[42,43] also support the interchange magnetic reconnection scenario. We note that structured microstreams and magnetic switchbacks are present throughout the inner heliosphere measured by the PSP and that the primary difference between the slow and fast solar wind may lie in the magnetic topology of the underlying coronal hole.

## Online content

Any methods, additional references, Nature Portfolio reporting summaries, source data, extended data, supplementary information, acknowledgements, peer review information; details of author contributions and competing interests; and statements of data and code availability are available at https://doi.org/10.1038/s41586-023-05955-3.

1. McKenzie, J. F., Banaszkiewicz, M. & Axford, W. I. Acceleration of the high speed solar wind. *Astron. Astrophys.* **303**, L45 (1995).
2. Axford, W. I. et al. Acceleration of the high speed solar wind in coronal holes. *Space Sci. Rev.* **98**, 25 (1999).
3. Fisk, L. A., Schwadron, N. A. & Zurbuchen, T. H. Acceleration of the fast solar wind by the emergence of new magnetic flux. *J. Geophys. Res.* **104**, 19765 (1999).
4. Cranmer, S. R. & van Ballegooijen, A. A. Can the solar wind be driven by magnetic reconnection in the Sun's magnetic carpet? *Astrophys. J.* **720**, 824 (2010).
5. Zank, G. P. et al. Theory and transport of nearly incompressible magnetohydrodynamic turbulence. IV. Solar coronal turbulence. *Astrophys. J.* **854**, 32 (2018).
6. Fox, N. J. et al. The Solar Probe Plus mission: humanity's first visit to our star. *Space Sci. Rev.* **204**, 7 (2016).
7. Bale, S. D. et al. Highly structure slow solar wind emerging from an equatorial coronal hole. *Nature* **576**, 237 (2019).
8. Kasper, J. C. et al. Alfvénic velocity spikes and rotational flows in the near-Sun solar wind. *Nature* **576**, 228 (2019).
9. Bale, S. D. et al. A solar source of Alfvénic magnetic field switchbacks: in situ remnants of magnetic funnels on supergranulation scales. *Astrophys. J.* **923**, 174 (2021).
10. Rieutord, M. & Rincon, F. The Sun's supergranulation. *Living Rev. Sol. Phys.* **7**, 2 (2010).
11. Kasper, J. C. et al. Solar Wind Electrons Alphas and Protons (SWEAP) investigation: design of the solar wind and coronal plasma instrument suite for Solar Probe Plus. *Space Sci. Rev.* **204**, 131 (2016).
12. McComas, D. J. et al. Integrated Science Investigation of the Sun (ISIS): design of the energetic particle investigation. *Space Sci. Rev.* **204**, 187 (2016).
13. Bale, S. D. et al. The FIELDS instrument suite for Solar Probe Plus: measuring the coronal plasma and magnetic field, plasma waves and turbulence, and radio signatures of solar transients. *Space Sci. Rev.* **204**, 49 (2016).
14. Thieme, K. M., Marsch, E. & Schwenn, R. Spatial structures in high-speed streams as signatures of fine structures in coronal holes. *Ann. Geophys.* **8**, 713 (1990).
15. Neugebauer, M. et al. Ulysses observations of microstreams in the solar wind from coronal holes. *J. Geophys. Res.* **100**, 23389 (1995).
16. Kasper, J. C. et al. Solar wind helium abundance as a function of speed and heliographic latitude: variation through a solar cycle. *Astrophys. J.* **660**, 901 (2007).
17. Schatten, K. H., Wilcox, J. M. & Ness, N. F. A model of interplanetary and coronal magnetic fields. *Solar Phys.* **6**, 442 (1969).
18. Altschuler, M. D. & Newkirk, G. Magnetic fields and the structure of the solar corona. 1: methods of calculating coronal fields. *Solar Phys.* **9**, 131 (1969).
19. Hoeksema, J. T. *Structure and Evolution of the Large Scale Solar and Heliospheric Magnetic Fields*. PhD thesis, Stanford Univ. (1984).
20. Lemen, J. R. et al. The Atmospheric Imaging Assembly (AIA) on the Solar Dynamics Observatory (SDO). *Solar Phys.* **275**, 17 (2012).
21. Fisk, L. A. & Kasper, J. C. Global circulation of the open magnetic flux of the Sun. *Astrophys. J. Lett.* **894**, L4 (2020).
22. Zank, G. P. et al. The origin of switchbacks in the solar corona: Linear theory. *Astrophys. J.* **903**, 1 (2020).
23. Drake, J. F. et al. Switchbacks as signatures of magnetic flux ropes generated by interchange reconnection in the corona. *Astron. Astrophys.* **650**, A2 (2021).
24. Drake, J. F. et al. Formation of secondary islands during magnetic reconnection. *Geophys. Res. Lett.* **33**, 13015 (2006).
25. Bhattacharjee, A. et al. Fast reconnection in high-Lundquist-number plasmas due to the plasmoid instability. *Phys. Plasmas* **16**, 112102 (2009).
26. Cassak, P. A., Shay, M. A. & Drake, J. F. Scaling of Sweet–Parker reconnection with secondary islands. *Phys. Plasmas* **16**, 120702 (2009).
27. Daughton, W. et al. Transition from collisional to kinetic regimes in large-scale reconnection layers. *Phys. Rev. Lett.* **103**, 065004 (2009).
28. Drake, J. F. et al. Ion heating resulting from pickup in magnetic reconnection exhausts. *J. Geophys. Res.* **114**, A05111 (2009).
29. Drake, J. F., Swisdak, M. & Fermo, R. The power-law spectra of energetic particles during multi-island magnetic reconnection. *Astrophys. J. Lett.* **763**, L5 (2013).
30. Zhang, Q., Guao, F. & Daughton, W. Efficient nonthermal ion and electron acceleration enabled by the flux-rope kink instability in 3D nonrelativistic magnetic reconnection. *Phys. Rev. Lett.* **127**, 185101 (2021).
31. Cranmer, S. R. et al. An empirical model of a polar coronal hole at solar minimum. *Astrophys. J.* **511**, 481 (1999).
32. Shay, M. A. et al. The scaling of collisionless magnetic reconnection for large systems. *Geophys. Res. Lett.* **26**, 2163 (1999).
33. Shay, M. A., Drake, J. F. & Swisdak, M. Two-scale structure of the electron dissipation region during collisionless magnetic reconnection. *Phys. Rev. Lett.* **99**, 155022 (2007).
34. Daughton, W. et al. Computing the reconnection rate in turbulent kinetic layers by using electron mixing to identify topology. *Phys. Plasmas* **21**, 052307 (2014).
35. Huang, Y.-M. & Bhattacharjee, A. Turbulent magnetohydrodynamic reconnection mediated by the plasmoid instability. *Astrophys. J.* **818**, 20 (2016).
36. Burch, J. L. et al. Plasma injection and transport in the mid-latitude polar cusp. *Geophys. Res. Lett.* **9**, 921 (1982).
37. Withbroe, G. L. The temperature structure, mass, and energy flow in the corona and inner solar wind. *Astrophys. J.* **325**, 442 (1988).
38. Parker, E. N. Dynamics of interplanetary gas and magnetic fields. *Astrophys. J.* **128**, 664 (1958).
39. Axford, W. I. & McKenzie, J. F. The origin of high speed solar wind streams. In *Solar Wind Seven* (eds Marsch, E. & Schwenn, R.) 1–5 (Pergamon Press, 1992).
40. Ergun, R. E. et al. Observations of particle acceleration in magnetic reconnection-driven turbulence. *Astrophys. J.* **898**, 154 (2020).
41. Phan, T. D. et al. Parker Solar Probe observations of solar wind energetic proton beams produced by magnetic reconnection in the near-Sun heliospheric current sheet. *Geophys. Res. Lett.* **49**, e96986 (2022).
42. Telloni, D. et al. Observation of a magnetic switchback in the solar corona. *Astrophys. J. Lett.* **936**, L25 (2022).
43. Raouafi, N. E. et al. Magnetic reconnection as the driver of the solar wind. *Astrophys. J.* **945**, 28 (2023).

## Methods

### Potential field source surface modelling

To generate the footpoints shown in Figs. 1e and 2, a PFSS model[16,17] was run using an Air Force Data Assimilative Photospheric Flux Transport–Global Oscillation Network Group (ADAPT–GONG) magnetogram[44–46] from 21 November 2022, with a source surface height set to the canonical value of $2.5 R_S$[19] by means of the open-source pfsspy[47] software. The footpoint mapping from the PSP down to the solar surface followed the methodology[48] comprising a ballistic heliosphere[49,50] and the PFSS domain from $2.5 R_S$ down to the photosphere[51].

The results for PSP E10 were distinct and compelling. As shown in Figs. 1 and 2, for 20–21 November, the PSP was rotating faster than the Sun and moving from left to right in the Carrington frame of reference shown in those plots. The footpoint mapping connected deep inside two mid-latitude negative polarity coronal holes of substantial area. This source mapping is uniquely well supported, compared with previous PSP encounters, owing to the comparison of the in situ data. First, the magnetic polarity measured by the PSP throughout the encounter is well explained by the PFSS current sheet geometry and coronal hole polarity. Second, the times when the PSP maps to the centre of these large coronal holes correspond to maxima in solar wind speed, and at the time the connection switches from one source to another in the model, there is a distinct dip in solar wind speed, clearly consistent with the traversal of overexpanded field lines at the coronal hole boundaries[52]. This correspondence is clearly shown in Fig. 1 where the transition between 'stream 1' and 'stream 2' marked in the bottom panel corresponds to the dip in solar wind speed (black trace, Fig. 1c).

### PSP/SWEAP data analysis

We use proton and alpha particle measurements from the Solar Wind Electrons Alphas and Protons (SWEAP) instrument suite[11] on PSP. The proton spectrum in Fig. 4 is taken from the Solar Probe Analyzer (SPAN-Ion) SF00 data product, averaged over the time range from 20 November 2021 at 04:00:00 to 20 November 2021 at 19:00:00 and summed over all look directions. We work in units of energy flux as opposed to number flux or distribution function as it results in a spectrum spanning fewer orders of magnitude at high energy, facilitating the comparison between SPAN-Ion and Integrated Science Investigation of the Sun (ISOIS)/Epi-Lo data, as well as being the quantity most directly related to the SPAN-Ion measurements. The power law for the protons is fit to the four highest energy SPAN data points and the ISOIS data points. The alpha spectrum is obtained in the same way from SPAN-Ion's sf01 data product, except that a small amount (around 1%) of contaminant protons leaking in from the sf00 channel are accounted for and subtracted. The large shift to higher energy of the alphas relative to the protons during this interval means that the contaminant protons have no impact on the power law part of the spectrum or its exponent, and only affect the lowest energy data points.

### Estimation of the reconnection magnetic field

The strength of the magnetic field that drives interchange reconnection will control the rate of magnetic energy release and the spectra of energetic particles produced. Although the SDO/HMI observations make known the structure of the magnetic field in the low corona, these measurements do not show the strength of the magnetic field that is actually undergoing reconnection because there is substantial variation of the field strength along the surface. To estimate the strength of the magnetic field driving the flow bursts measured by the PSP, we project the measured magnetic field at the PSP and project this magnetic field down to the solar surface. The radial magnetic field $B_R$ at the perihelion of E10, as shown in Fig. 1, is around 600 nT. Direct measurements of the radial profile of $B_R$ over the first five PSP orbits have established an $R^{-2}$ scaling for the field[45], consistent with the conservation of the radial magnetic flux. However, deviations from this scaling are expected close to the Sun. Specifically, because closed flux occupies a substantial fraction of the solar surface, the open flux will be compressed into a reduced fraction of the solar surface, which will lead to greater compression of the magnetic field near the solar surface. A rough estimate of the increased magnetic field compression can be obtained by averaging the radial magnetic field obtained from the PFSS model during the E10 perihelion. The radial dependence of this averaged field is shown in Extended Data Fig. 1. The magnetic field compression from $2.5 R_s$, the outer boundary of the PFSS grid, down to just above the solar surface is around 26, which is well above the compression of around 6.25 from the $R^{-2}$ dependence. Thus, we assume that the $R^{-2}$ describes the radial dependence from $R = 13.4 R_s$ to $2.5 R_s$ and take the compression of 26 from $2.5 R_s$ to just above the solar surface. The projection of the 600-nT field down to the solar surface is around 4.5 G, which is in reasonable agreement with the strength of the solar surface magnetic field shown in Fig. 2.

### Particle-in-cell simulations

Our estimate of the rate of interchange reconnection based on projections of the PSP observations back to the low corona suggest that reconnection there is deeply in the collisionless regime. To explore the structure of the interchange reconnection exhaust and the resulting energetic proton and alpha spectra measured at the PSP, we use the PIC model p3d (ref. 53). The MHD model is not adequate to explore the particle energization documented in the PSP data. We limit the calculations to a two-dimensional system with an initial magnetic geometry that leads to reconnection between open and closed flux low in the corona[3,54]. Because of constraints on the domain size possible with the PIC model, there is no gravity in the simulations, so the model does not describe the complete dynamics of the solar wind drive mechanism. In addition, line-tied boundary conditions are not imposed at the nominal coronal surface. Thus, the model is not a complete description of interchange reconnection in the low corona, but will provide information on the dynamics of collisionless reconnection, the structure of the outflow exhaust, the bulk heating and the spectra of accelerated particles. We include alpha particles (5% by number) so that the spectra of protons and alphas can be compared.

The initial state for the simulation consists of a band of vertical flux (field strength $B_0$ in the negative radial direction) with a low plasma density ($0.1 n_0$) and an adjacent region with higher density that is a cylindrical equilibrium. The detailed initial state has been described previously[23], so the governing equations are not repeated here. The peak magnetic field of the cylindrical equilibrium is $0.76 B_0$ with a peak density of $n_0$. The temperatures are uniform with $T_e = T_p = T_a = 0.06 m_p V_{A0}^2$ with $V_{A0}$ the Alfvén speed based on $B_0$, $n_0$ and the proton mass $m_p$. Thus, in the initial state, the plasma pressure is small compared with the magnetic pressure, as expected in the corona. The guide field $B_z$ is non-zero everywhere with a profile that balances the pressure and tension forces. The strength of the guide field can be varied by choosing its value in the region of vertical flux. Its value does not substantially affect the overall structure and dynamics of reconnection shown in Fig. 3. However, because a strong guide field weakens the Fermi drive mechanism for particle energy gain, the guide field controls the power-law index of energetic protons and alphas. The energy fluxes shown in Fig. 4 were from a simulation with a guide field of $0.55 B_0$. Simulations with a weaker (stronger) guide field produced harder (softer) spectra.

The results of the simulation are presented in normalized units: times to the Alfvén transit time across the domain of scale length $L$, $L/V_{A0}$, plasma fluxes to $n_0 V_{A0}$ and energies to $m_p V_{A0}^2$. The domain dimensions in the x and y directions are equal. The mass ratio $m_p/m_e = 25$ is artificial as is the velocity of light ($20 V_{A0}$) and the proton inertial scale $d_p = L/163.84$. As has been established in earlier papers, the results are not sensitive to these values[33,34]. The radius of the cylindrical magnetic field is $60 d_p$ and the grid scales are $0.02 d_p$ in both space directions, with around 400 particles per cell.

# Article

Because the velocities and energies in the simulation are normalized to the Alfvén speeds $V_{A0}$ and $m_p V_{A0}^2$, respectively, direct comparison with the observations requires that these parameters, and specifically the Alfvén speed $V_{A0}$ where reconnection is taking place, be established. As described in the main text, we use two distinct approaches for estimating $V_{A0}$. The first comes from the amplitude of the flow bursts measured by the PSP at 13.4 $R_s$, which have values of around 300 km s$^{-1}$. The second comes from comparing the proton spectra from the simulation and that measured with SPAN-Ion. Specifically, we equate the low energy limit of the proton power-law spectrum from the simulation and the observations. This yields $m_p V_{A0}^2 = 1.4$ keV, which corresponds to $V_{A0} = 370$ km s$^{-1}$. Thus, the two approaches yield comparable values, which enables us to directly compare the simulation results with the observations. The spectral indices of around −8 for the proton and alpha energy fluxes from the simulation are independent of this normalization. That they are in approximate agreement with the observational data is strong support for the interchange reconnection model for these energetic particles. We further note that a simulation with half of the domain size ($L = 81.92\, d_p$) produced power-law spectra with similar spectral indices.

Many of the global models that have been used to explore the acceleration of the solar wind have been based on the assumption that Alfvén waves[1,2] or other forms of magnetic structures[5] are injected into the low corona and the heating associated with this turbulence produces the pressure required to drive the wind. As interchange reconnection is often invoked as the source of this turbulence, it is important to explore whether a substantial fraction of released magnetic energy appears as magnetic turbulence versus direct particle energization or bulk flow. Shown in Extended Data Fig. 2 is a two-dimensional plot of the proton temperature from the same simulation and at the same time as the plot of the vertical proton flux in Fig. 3b. The entire outflow exhaust is filled with high temperature protons with temperatures that are a large fraction of $m_p V_{A0}^2$, which, as discussed in the main text, is in the range of 0.9 keV to 1.4 keV. The warped field lines in the reconnection exhaust that are evident in Fig. 3b and Extended Data Fig. 2 suggest that reconnection also drives waves and magnetic turbulence. A detailed exploration of the relative energy that appears as magnetic turbulence versus direct heating has not been carried out but is an important extension of the present results, especially in three dimensions, where the process of reconnection is much more dynamic.

## Data availability

The PSP mission data used in this study are openly available at the NASA Space Physics Data Facility (https://nssdc.gsfc.nasa.gov) and were analysed using the IDL/SPEDAS software package (https://spedas.org/blog/). The computer simulations used resources of the National Energy Research Scientific Computing Center, a DOE Office of Science User Facility supported by the Office of Science of the US Department of Energy under contract no. DE-AC02-05CH11231. Simulation data is available at https://doi.org/10.5281/zenodo.7562035.

44. Arge, C. N. et al. Air Force Data Assimilative Photospheric Flux Transport (ADAPT) model. In *AIP Conferenece Proceedings* (eds Maksimovic, M. et al) **1216**, 343 (AIP, 2010).
45. Arge, C. N. et al. Improving data drivers for coronal and solar wind models. In *5th International Conference of Numerical Modeling of Space Plasma Flows (ASTRONUM 2010)* (eds Pogorelov, N. V. et al) **444**, 99–104 (ASP, 2011).
46. Arge, C. N. et al. Modeling the corona and solar wind using ADAPT maps that include far-side observations. In *AIP Conference Proceedings* **1539**, 11–14 (AIP Publishing LLC, 2013).
47. Stansby, D., Yeates, A. & Badman, S. pfsspy: a Python package for potential source surface modeling. *J. Open Source Software* **5**, 2732 (2020).
48. Badman, S. T. et al. Magnetic connectivity of the ecliptic plane within 0.5 au: potential field source surface modeling of the first Parker Solar Probe encounter. *Astrophys. J. Supp.* **246**, 23 (2020).
49. Nolte, J. T. & Roelof, E. C. Large-scale structure of the interplanetary medium, I: high coronal source longitude of the quiet-time solar wind. *Solar Phys.* **33**, 241 (1973).
50. MacNeil, A. R. et al. A statistical evaluation of ballistic backmapping for the slow solar wind: the interplay of solar wind acceleration and corotation. *MNRAS* **10**, 1093 (2021).
51. Badman, S. T. et al. Prediction and verification of Parker Solar Probe solar wind sources at 13.3 $R_s$. *J. Geophys. Res.* **128**, e2023JA031359 (2023).
52. Wang, Y.-M. & Sheeley, N. R. Jr Solar wind speed and coronal flux-tube expansion. *Astrophys. J.* **355**, 726 (1990).
53. Zeiler, A. et al. Three-dimensional particle simulations of collisionless magnetic reconnection. *J. Geophys. Res.* **107**, 1230 (2002).
54. Fisk, L. A. The open magnetic flux of the Sun. I. Transport by reconnections with coronal loops. *Astrophys. J.* **636**, 563 (2005).

**Acknowledgements** The FIELDS, SWEAP and ISOIS suites were designed, developed and are operated under NASA contract NNN06AA01C. We acknowledge the extraordinary contributions of the PSP mission operations and spacecraft engineering team at the Johns Hopkins University Applied Physics Laboratory. M.V. was supported in part by the International Space Science Institute, Bern, through the J. Geiss fellowship. J.F.D. and M.S. were supported by the NASA Drive Science Center on Solar Flare Energy Release (SolFER) under Grant 80NSSC20K0627, NASA Grant 80NSSC22K0433 and NSF Grant PHY2109083. T.S.H. is supported by STFC grant ST/W001071/1. O.P. was supported by the NASA Grant 80NSSC20K1829. Elements of this work benefited from discussions at the meeting of Team 463 at the International Space Science Institute (ISSI).

**Author contributions** S.D.B. and J.F.D. wrote the manuscript with major contributions from S.T.B. S.D.B. analysed the PSP measurements, with contributions from M.D.M., M.I.D., T.S.H. and D.E.L. S.T.B. performed the PFSS analysis. J.F.D. and M.S. performed the computer simulations. S.D.B., J.C.K. and D.J.M. lead the PSP/FIELDS, SWEAP and ISOIS teams, respectively. All authors participated in the data interpretation and read and commented on the manuscript.

**Competing interests** The authors declare no competing interests.

**Additional information**
**Correspondence and requests for materials** should be addressed to S. D. Bale.
**Peer review information** *Nature* thanks Vadim Uritsky, G. P. Zank and the other, anonymous, reviewer(s) for their contribution to the peer review of this work.
**Reprints and permissions information** is available at http://www.nature.com/reprints.

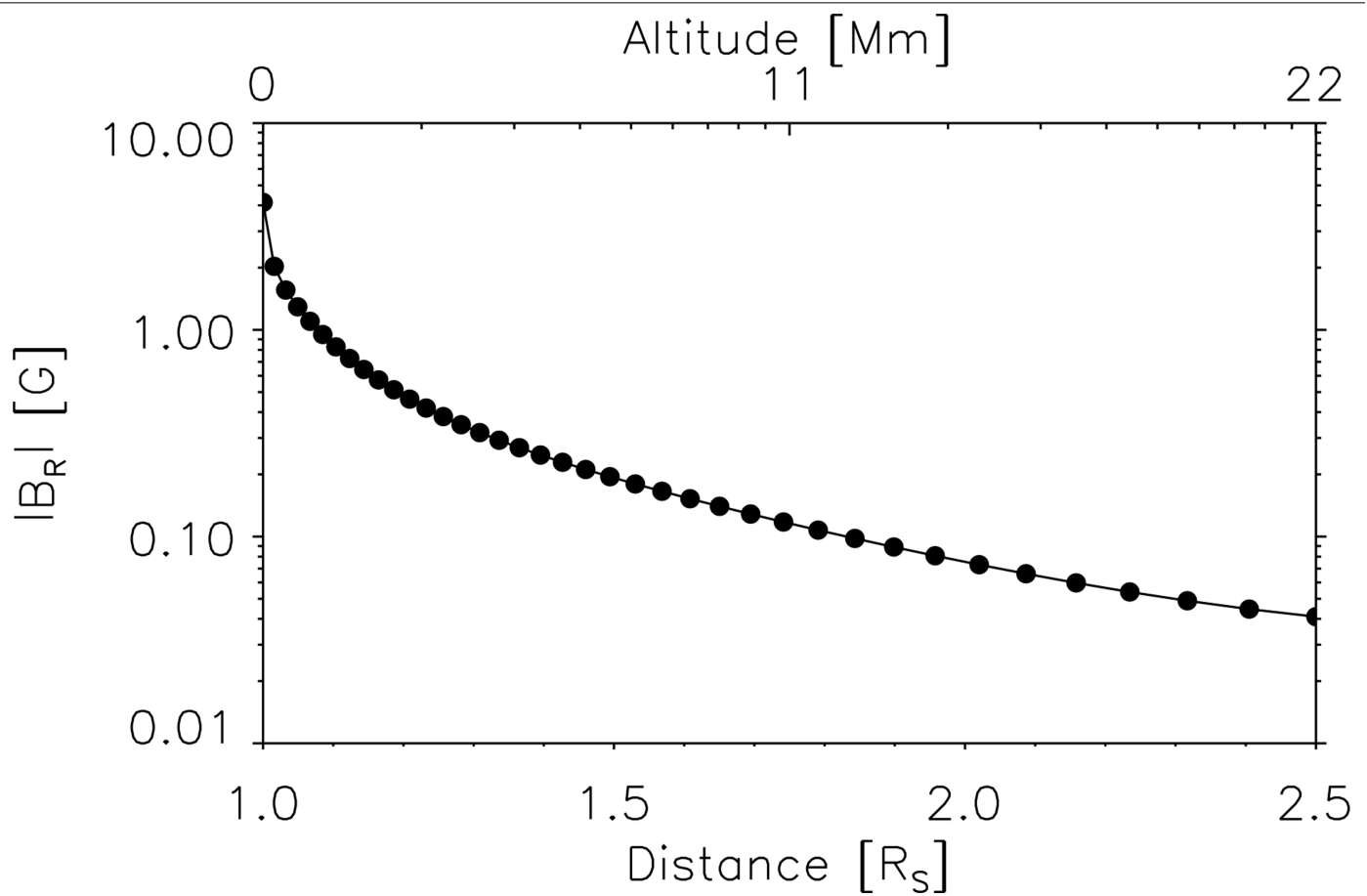

**Extended Data Fig. 1 | Magnitude of the radial magnetic field from PFSS.** The magnitude of the radial component of the magnetic field is modelled by the PFSS implementation, constrained by magnetograms at the photospheric footpoints and the open boundary condition at $2.5\,R_S$. This field magnitude is consistent with the field measured at Parker Solar Probe and is used to estimate the Alfvén speed at the reconnection site.

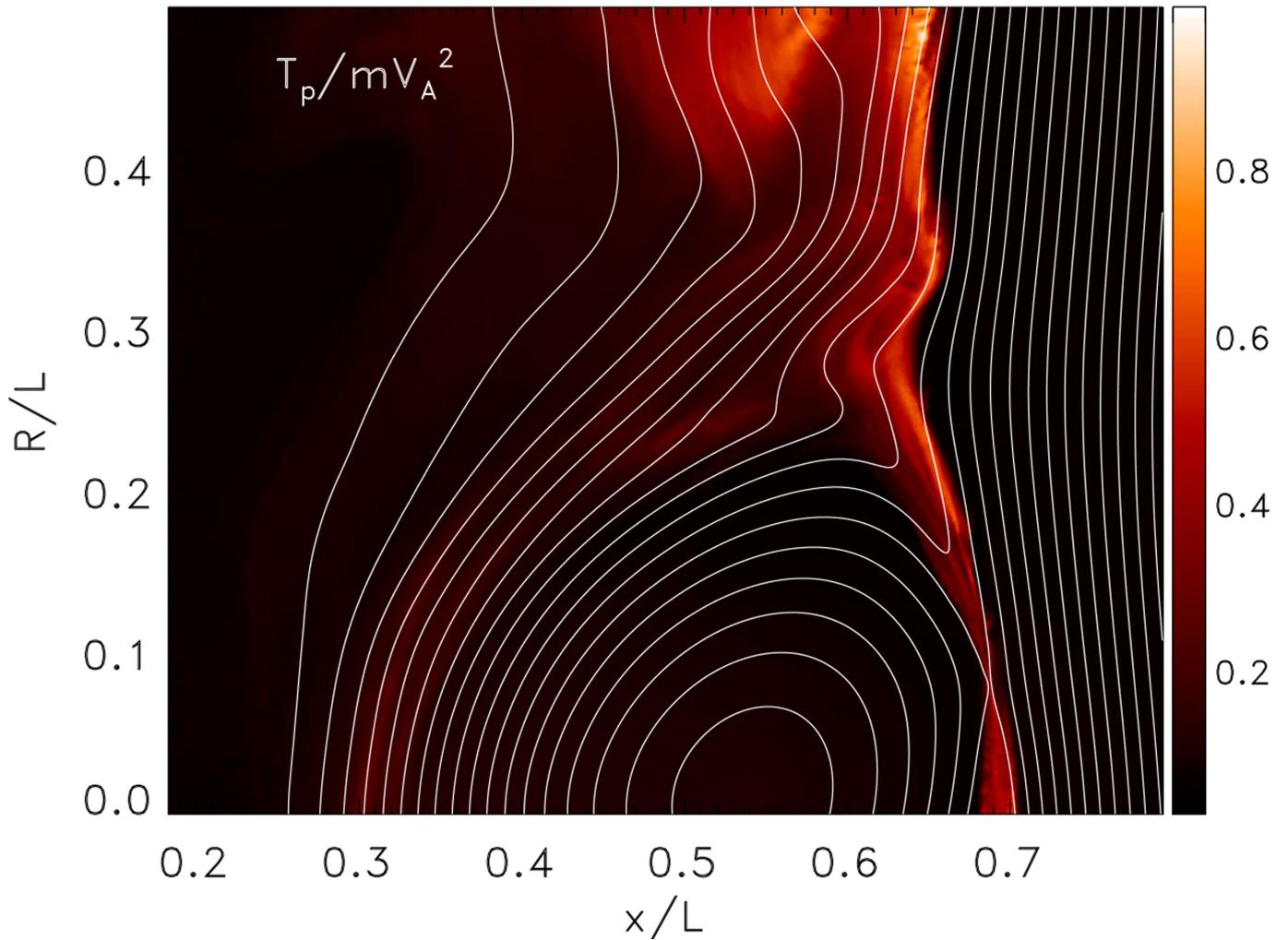

**Extended Data Fig. 2 | Proton temperature in the plane of reconnection with overlying magnetic field lines.** The proton temperature is shown from the same simulation and the same time as the vertical flux shown in Fig. 3b. The temperature is normalized to $m_p V_{A0}^2$, which, as discussed in the main text is in the range of 0.9 keV to 1.6 keV. Thus, since the magnetic energy released per particle during reconnection is around $m_p V_{A0}^2$, a large fraction of the released magnetic energy is goes into the heating and energization of the ambient plasma very close to the magnetic energy release site. The energy going into turbulent magnetic fields that is injected upward into the corona and is available to further heat the ambient plasma remains to be explored.